\documentclass[aps,
 prb
 mph,
 amsmath,amssymb,
 reprint,%
]{revtex4-2}

\usepackage{graphicx}
\usepackage{dcolumn}
\usepackage{bm}
\usepackage{color}

\begin{document}

\preprint{AAPM/123-QED}

\title{Composition effect in the thermo-mechanical behavior of glasses, and its modelization}

\author{R. Alvarez-Donado$^{1,2}$}
\email{rene-alberto.alvarez-donado@insa-lyon.fr}
 \author{M. Sepulveda-Macias$^3$}%
\author{A. Tanguy$^{1,4}$}
\email{anne.tanguy@insa-lyon.fr}
\affiliation{$^1$ Univ Lyon, INSA-Lyon, CNRS, LaMCoS, UMR 5259, Villeurbanne, 69621, France.\\
$^2$ Laboratoire de Mécanique des Solides, CNRS, École Polytechnique, Route de Saclay, Palaiseau, 91128, France.\\
$^3$ Departamento de Física, Facultad de Ciencias, Universidad de Chile, Casilla 653, Santiago, Chile.\\
$^4$ ONERA, Université Paris-Saclay, Palaiseau, 92120, France.
}

\date{\today}

\begin{abstract}
We employed molecular dynamics simulations to explore comparatively the thermo-mechanical behavior of two glass materials—an oxide silica glass (SiO$2$) and a binary Cu-Zr-based metallic alloy (Cu$_{50}$Zr$_{50}$)—during shear \textcolor{black}{and elongation} deformation cycles.  By calculating the energy balance and tracking the temperature evolution of both glasses under deformation cycles,  we are able to propose, for each of them, a constitutive law which accurately reproduces the self-heating process due to plastic deformation.  These relatively simple constitutive laws involve strain rate sensitivity and a non-linear temperature dependence of the thermal dilatancy coefficients, as well as strain gradient plasticity. To identify the right parameters, both glasses are equilibrated at very low temperature (10 K) and two independent deformation rates were applied to each sample \textcolor{black}{for each type of deformation.} Thermal attenuation is greatly amplified in silica compared to the metallic glass. Moreover, using precise atomic description of the instantaneous deformation, combined with exact coarse-graining procedure, we show\textcolor{black}{, in silica,} that self-heating is mainly supported by inhomogeneous strain gradient plasticity with nanometric characteristic lengthscales.
\end{abstract}

\maketitle

\section{Introduction}
Glasses are solid materials typically obtained by sufficiently rapidly cooling a liquid to prevent crystallization. As a result, glasses inherit an amorphous internal structure close to that of their liquid phase, which endows them with exceptional properties. For example, metallic glasses (MG) have a large elastic limit strain and high yield strength\cite{Ashby2006,schroers2013bulk,taloni2018size} compared to highest resistant steels; oxyde glasses are harder than quartz, they have a high dielectric constant but low thermal conductivity and are used, for example, as fiber-reinforced polymer composites\cite{thyagarajan2007fiber,chowdhury2016molecular,mallick2007fiber}. However, unlike crystalline solids, a complete understanding of the elasto-plastic regime or the mechanisms underlying the temperature dependence of plasticity in glasses remains a topic of intense research. Albaret et al.\cite{albaret2016mapping} have shown that in the quasistatic athermal limit, plastic rearrangements can be described as a athermal succession of Eshelby-like local irreversible events\cite{albaret2016mapping}. Additionally, Beltukov et al.~\cite{beltukov2016boson,beltukov2018transport,tanguy2023vibrations} have observed that, due to scattering processes in amorphous structures, mechanical wave packets can induce local heating and heat transfer in glassy materials even at very low temperature. Moreover, the mechanical behaviour of glasses, especially its plastic behaviour, is sensitive not only to the surrounding temperature, but also to the strain rate\cite{varnik2006temperature}, thus raising the question of the time-temperature superposition~\cite{berthierbarrat2002,pelletier2014}. Likewise, the combined frequency and temperature dependence of the thermo-mechanical response of glasses suggests an intimate relationship between thermal and mechanical responses\cite{tsamados2010,fusco2010role,fusco2014,pelletier2024,sepulveda2024thermomechanical}.\\
\indent Since temperature changes inside glass can compromise its mechanical response, such as strength, softening, or ductility\cite{pelletier2014}, thermomechanical couplings must be taken into account when determining constitutive laws. However, attempts to study the conversion of plastic work into heat during large strain deformation, whether experimentally or through numerical simulations, have yielded results that strongly depend on the material and its composition\cite{ravichandran2002conversion,kositski2021employing,xiong2022atomistic,signetti2023quantification}. The local temperature rise in glasses is directly related to the formation of shear bands (SB)\cite{battezzati2008quantitative,miracle2011shear,kato1997synthesis,conner1997fracture}. In fact, the zones that experience the largest irreversible strains, i.e., the SB, contribute the most to self-heating. Unfortunately, measuring the temperature inside the SB is challenging, and \textcolor{black}{the experimental magnitude} of local temperature increase remains highly controversial, with estimates ranging from nearly 0.1 K to around 1000 K\cite{bengus1993some,liu1998test,yang2004situ,yang2005dynamic,bruck1996dynamic,flores1999local,lewandowski2006temperature,gilbert1999light,wright2001localized}. Factors such as the thickness and duration of the SB should also be considered when determining self-heating. As pointed out by Lewandowski and Greer\cite{lewandowski2006temperature}, when thermal diffusion and the local shear rate are correctly taken into account, predictions of the temperature rise vary from 40 K to thousands of kelvins\cite{lewandowski2006temperature,bengus1993some}. Thus, there is no consistent estimate of the temperature rise in glasses. Therefore, in this work, we propose a constitutive law that takes the aforementioned parameters into account to correctly describe self-heating in amorphous materials.\\
\indent Regarding theoretical approaches, in a seminal work, Chrysochoos et al. \cite{chrysochoos1992thermographic} established thermo-mechanical constitutive laws through the generalized standard material theory to describe the energy balance evolution of four different materials (duralumin, brass, carbon, and stainless steel) during cycles of tensile deformation. They observed that the model based on the classical theory of time-independent elastoplasticity provides accurate mechanical predictions. Zhao and Li\cite{zhao2011local} used finite element simulations to solve the heat equation, using the work accumulated during plastic deformation as a heat source. Lagogianni and Varnik\cite{lagogianni2022temperature}, performing molecular dynamics simulations, showed that the local temperature rise occurs primarily within the shear band compared to the matrix outside it. Following these lines, in this work we propose a fully coupled thermo-mechanical constitutive law able to describe the temperature rise in glasses, taking into account the shear rate, the thermal expansion coefficients, the thermomechanical couplings, as well as the local atomic-scale deformation gradient during deformation cycles, as suggested within the framework of strain gradient plasticity\cite{GradientPlasticity1992}. In order to test the universality of this behaviour in glasses, classical molecular dynamics (MD) simulations are used to compute the stress-strain shear deformation curves of two structurally different families of glasses: a silica (SiO$_2$) oxide glass and a metallic alloy of CuZr (Cu$_{50}$Zr$_{50}$), submitted to cyclic deformations with two different deformation rates. While oxide glasses range from covalent materials with open structures to mixed oxides with ionic bonding, metallic glasses are essentially close-packed, with bonding dominated by delocalized metallic orbitals\cite{gaskell1982local}. In this sense this work can be considered as an extension of our previous one that proposed a constitutive law for metallic glasses at different strain rates and temperatures\cite{sepulveda2024thermomechanical}. Thus, the validity of our macroscopic thermo-mechanical model at different strain rates and temperatures is demonstrated by how well the temperature rise is described in these different families of glasses.\\
\indent The paper is organized as follows: Section II presents our simulation setup and the protocol followed during our simulations. Section III describes the constitutive model and the results of our simulations. Finally, Section IV provides our conclusions and future perspectives.

\section{Simulation setup}
\begin{figure}[htb]
\includegraphics{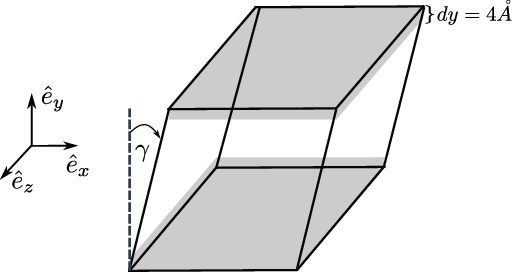}
\caption{Schematic representation of the deformation protocol. The grey layers represent the regions of atoms where the temperature is fixed by a thermostat during an xy-shear deformation. The rest of the atoms are kept in the NVE ensemble and used to track the temperature evolution during the deformation cycles.}
\label{fig_1}
\end{figure}

The interatomic interactions for the CuZr-based alloys were described by a modified embedded atom method (MEAM)\cite{baskes1992modified,kim2008modified}, while for the case of SiO$_2$, we modeled the interactions with an empirical BKS potential\cite{van1990force} using the parameters of Yuan and Cormack\cite{yuan2001local} with a cut-off distance of 4.8 $\text{\AA}$. We focused on two system sizes: a small one with a cubic box of $L\approx 50\text{\AA}$ , and a medium system in a rectangular box with dimensions $L_x = L_y \approx 150\text{\AA} $  and $L_z \approx 50\text{\AA}$. In both cases we impose periodic boundary conditions in all three dimensions. Our simulations were carried out in the classical MD open code LAMMPS\cite{thompson2022lammps} with a timestep $\Delta t = 1fs$, while the structure visualization
and the color coding are done by OVITO\cite{stukowski2009visualization}. To begin with, we randomly distributed atoms in the simulation box and immediately performed a minimization to prevent any possible atom overlap. Since the glass transition temperature differs for every material, SiO$_2$ was quenched from 6000 K to 10 K, while the CuZr alloy was quenched from 2000 K to 10 K. In both cases, the time step $\Delta t$ was $1 fs$, with a cooling rate of 100 K$/ns$, and we maintained the temperature and pressure using the Nosé-Hoover thermostat and barostat with damping parameters of $\tau_T= 100 fs$ and $\tau_p = 5 ps$ , respectively. All results were obtained while maintaining an external pressure $P = 0$ Bars. The equilibrated samples transitioned to the $NVT$ ensemble and the simulation box was rescaled to the average equilibrium volume obtained in the $NPT$ ensemble, resulting in a mean external pressure $\langle P \rangle = 0$ Bars. Once we prepared the glasses in the solid state, we performed a series of shear deformation cycles using two constant deformation rates, $\dot{\gamma} = 10^{9} \text{s}^{-1}$ and $10^{10} \text{s}^{-1}$ that correspond to relatively slow strain rates in MD, not far from the quasi-static case. Note that such high strain rate values look far from the one reachable experimentally, except in extreme cases of laser impacts\cite{raffray2023zr}. However, the correspondence to quantitative values is far from being established. The behaviour of materials depends on the realism in the description of the energy barriers, which depend on the empirical description used for the interatomic interactions, as will be discussed later. Such high values are thus not so far from the quasi-static case, as already noticed in\cite{fusco2014}. In any case, it is difficult to reach smaller strain rates values with classical molecular dynamics simulations. The two strain rate values used here will mainly allow to discuss possible strain rate sensitivities of the parameters in the constitutive laws. We track the temperature evolution of the glasses and its dependence on $\dot{\gamma}$ and size. When performing MD simulations, the temperature is controlled by incorporating a thermostat, thus attention must be paid to the shear and heating protocols. In our case, we apply a shear strain $\gamma = \gamma_{xy}$, generating deformation by tilting the top and bottom $xz$-planes. Moreover, we thermostat two layers of the system in this plane with a thickness of $4 \text{\AA}$, and use the rest of the atoms, which are kept in the $NVE$ ensemble, to compute the temperature evolution. Fig.~\ref{fig_1} depicts the schematic representation of the simulation setup employed during the deformation cycles.

\section{Theoretical Framework}
\begin{figure*}
\includegraphics{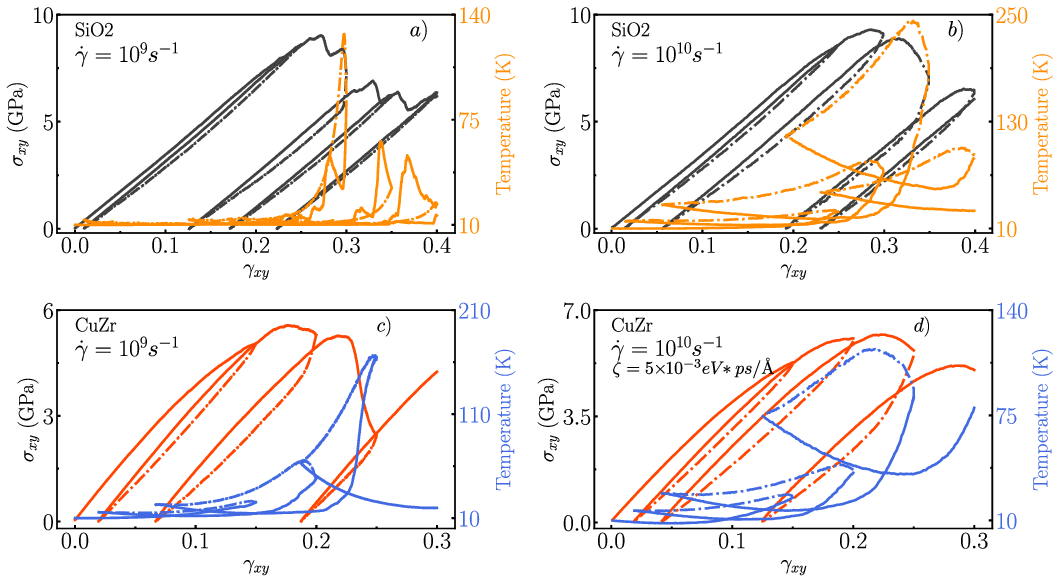}
\caption{Averaged thermo-mechanical responses of the medium size samples during the load and unload shear deformation of SiO$_2$ at $a)\:\:\:\dot{\gamma} = 10^9s^{-1}$ and $b)\:\:\:\dot{\gamma} = 10^{10}s^{-1}$. The values of temperature evolution in both cases are shown in the right axis in yellow. Similarly, the thermo-mechanical response of the medium size sample of CuZr at $c)\:\:\:\dot{\gamma} = 10^9s^{-1}$ and $d)\:\:\:\dot{\gamma} = 10^{10}s^{-1}$ with a damping parameter of $\zeta=5.10^{-3} eV.ps/\AA^2$ in the last case, as discussed in the text. The values of temperature evolution in both cases are shown in the right axis in blue. \textcolor{black}{Reverse loading is represented by dashed-dot lines.}}
\label{fig_2}
\end{figure*}
The deformation curves of the two medium samples at two different shear rates are depicted in Fig. \ref{fig_2}. For each stress-strain curve, we averaged the stress and temperature over eight independent simulations. The temperature profile is computed as the average of the kinetic energy fluctuations of the atoms outside the fixed temperature layers, after removing the streaming velocity and the rigid body motion of the mobile atoms. Fig \ref{fig_2}$a)$ and Fig \ref{fig_2}$b)$ show the behavior of SiO$_2$ at $\dot{\gamma} = 10^{9} \text{s}^{-1}$ and $10^{10} \text{s}^{-1}$, respectively. These strain rates are those accessible in MD simulations. They tend to reproduce the quasi-static response of the glass in the slowest case. Fig 2c) and 2d) show the same quantities at the same different strain rates, but for CuZr. An additional damping parameter is applied to obtain the deformation curves in Fig 2d), for the reasons that will be explained later. In both cases, the system exhibits an increase in temperature when plastic events occur, with the maximum temperature peak happening when the glass experiences the stress overshoot. For $\dot{\gamma} = 10^{9} \text{s}^{-1}$, the temperature profile remains almost constant until a deformation of approximately $25\%$, when the stress overshoot occurs and the temperature rises by more than hundred kelvin. For the fastest shear rate ($\dot{\gamma} = 10^{10} \text{s}^{-1}$), the temperature increases with every deformation step, reaching its maximum increase when the stress overshoot occurs at about $\gamma\approx 30\%$ strain in SiO$_2$, and only $\gamma\approx 20\%$ strain in CuZr, with a maximum temperature rise of $\Delta T \approx 250$ K in SiO$_2$, and only $\Delta T\approx 110$ K in CuZr. As expected, the stress overshoot strongly depends on the imposed deformation rate. 
In SiO$_2$, in the slower shear rate case, the glass goes back to its initial temperature when the stress goes back to zero. It is not the case when the shear rate is increased. For $10^{10}$ s$^{-1}$, the stress release is not sufficient to suppress  the temperature rise. The temperature thus continues decaying even when the strain increases again (Fig.~\ref{fig_2}). As will be seen \textcolor{black}{in Fig.~\ref{fig_4},} this effect is related in particular to the energy stored in the thermal dilatancy part of the elastic free energy. Characteristic butterfly shapes are then visible in the temperature evolution. The same kind of temperature dependence occurs in CuZr for all the strain rates. In case of SiO$_2$ we will see later that thermal dilatancy is not sufficient however to explain quantitatively this temperature decay.
Once again, for CuZr at the smallest studied strain rate $\dot\gamma=10^9$ s$^{-1}$, the maximum temperature rise is recorded at the stress overshoot at $\gamma = 25\%$ with a temperature difference of $\Delta T \approx 150$ K (Fig. \ref{fig_2}$c)$). But for $\dot{\gamma} = 10^{10} \text{s}^{-1}$, we encountered the same problem as reported in Ref~\cite{sepulveda2024thermomechanical} for the metallic glass, where the temperature increases monotonically with deformation, i.e., we do not see any significant temperature reduction when the deformation is removed, due to the difficulty to evacuate heat in the sample. This clearly unphysical result did not occur in the small sample of CuZr when deformed at the same shear rate, thanks to the dissipation at the interfaces, that are direclty connected to the thermostat. Therefore, we use the behavior of the small sample as a gauge for the physical behavior of the temperature profile of CuZr at the fastest deformation rate and we add a constant damping parameter. The temperature profile depicted in Fig. \ref{fig_2}$d)$ is thus obtained by imposing a global damping force $\vec{F}_i = -\zeta \vec{v}_i$ on every atom in the system. This additional damping force is a way to reproduce the underlying quantum dissipation sources not taken into account in the classical simulations, like the one induced by the Fermi rule related to electronic excitations/desexcitations or that due to bond breaking\cite{bertsch1983damping}. The value of $\zeta = 5\times 10^{-3} eV\:ps/\text{\AA}^2$ is carefully chosen to ensure that the temperature profile resembles the one obtained in the small sample. Interestingly, this value for the damping rate is very close to that due to the phonon-electron coupling in metals as observed in~\cite{Mason2015}. A comparison about the behavior of the medium sample without the damping force as well as a comparison with the small sample behavior at the same deformation rate can be found in reference\cite{sepulveda2024thermomechanical}. It is also important to make a cautionary remark regarding the shear rates employed in our simulations. Since in molecular dynamics the timestep is restricted at femtoseconds, common MD simulations are performed at maximum hundred of nanoseconds which impose a limitation in both the cooling and deformation rates reachables by simulations. However, deformations rates of $\dot{\gamma} = 10^{8}s^{-1}$ has already been considered representative of quasistatic deformation\cite{fusco2014}. Our guess is that the fact of employing empirical potentials to describe the interatomic interactions, could underestimating the amplitude of energy barriers, such that the dynamics corresponding to the low numerical time scale corresponds to that of larger experimental effective time scales. And thus the effective strain rate would be far smaller than the apparent one. In general, one must always be very careful about the quantitative interpretation of MD results. In this paper, two different strain rates are used to test the sensitivity of the parameters in the strain rate (viscous behavior)\\

\indent To derive a general thermo-mechanical constitutive law, we propose a continuum model, similar to \cite{chrysochoos1992thermographic}, elaborated where the glass is described as a homogeneous, isotropic, and linear thermoelastic solid with deformation energy density given by\cite{sadd2009chapter}:
\begin{equation}
\psi_{mec} = \frac{1}{2}\sigma_{ij}\varepsilon^{\sigma}_{ij},
\label{density}
\end{equation} 
where $\sigma$ is the stress tensor, $\sigma_{ij}$ are the stress tensor components and $\varepsilon^{\sigma}_{ij}$ are the temperature independent components to the strain tensor. Since we are considering the glass as a linear and homogeneous solid we have $\sigma_{ij} = C_{ijkl}\varepsilon^{\sigma}_{kl}$ and $\varepsilon^{\sigma}_{ij} = \varepsilon_{ij} - \varepsilon_{ij}^{T}$, where $\varepsilon_{ij}^{T} = \alpha_{ij}(T-T_0)$ is the temperature dependent part of the strain, at the scale of the box size with $\alpha_{ij}$ the coefficients of linear expansion, $T$ and $T_0$ the current and the inital temperatures respectively, $\varepsilon_{ij}$ the components of the total strain tensor, and $C_{ijkl}$ are the elastic moduli. Thus, Eq. (\ref{density}) is a second order polynomial in $(T-T_0)$:
\begin{eqnarray}
\psi_{mec} = \frac{1}{2}C_{ijkl}\varepsilon_{kl}\varepsilon_{ij} - &C_{ijkl}&\alpha_{kl}\varepsilon_{ij}(T-T_0) \nonumber\\&+& \frac{1}{2}C_{ijkl}\alpha_{kl}\alpha_{ij}(T-T_0)^2\label{mec-gen}
\end{eqnarray}
Considering the case where $\gamma = 2\varepsilon_{xy}$ is the only non-zero component of the strain tensor, and assuming small strain and local thermodynamical equilibrium, we can simplify Eq. (\ref{mec-gen}) and write the general form of the free energy, including the deformation energy density and the entropy contribution, as follows: 
\begin{equation}
\psi(\gamma,T) = \frac{1}{2}\mu\gamma^2-2\gamma\mu\alpha_{xy}(T-T_0) - \beta(T-T_0)^2 ,
\label{mech}
\end{equation} 
where $\mu$ is the shear modulus, $\beta$ is a parameter, and the linear expansion coefficient $\alpha_{xy}= A_0 + A_1 (T-T_0)$ may supposed to have a linear dependence on $T$. From a thermodynamic perspective, we have:
\begin{equation}
\psi = e - Ts,
\label{ther}
\end{equation}
where $e$ and $s$ are the energy and entropy per unit volume, respectively. Thus, the total differential of $\psi$ (Eqs (\ref{mech}) and (\ref{ther})) we have:
\begin{equation}
d\psi =de - Tds - sdT = \frac{\partial\psi}{\partial\gamma}\Bigr|_T d\gamma + \frac{\partial\psi}{\partial T}\Bigr|_\gamma dT\label{dther}
\end{equation}
Recalling that $\frac{\partial\psi}{\partial T}\vert_\gamma = -s$, multiplying both sides by the density of the glass $\rho$ and differentiating with respect to time, we obtain:
\begin{equation}
\rho\dot{e} = \rho T\dot{s} + \rho\frac{\partial \psi}{\partial\gamma}\Bigr|_T\dot{\gamma},
\label{rho}
\end{equation}
The first term in the RHS of Eq (\ref{rho}) can be written in a more suitable form using  $\dot{s}=-\frac{d}{dt}\left(\frac{\partial\psi}{\partial T}\Bigr|_\gamma\right)$. Derivating as a function of the two variables $T$ and $\gamma$ yields:
\begin{equation}
\rho T\dot{s} =-\rho T \frac{d}{dt}\left(\frac{\partial\psi}{\partial T}\Bigr|_\gamma\right) =  \rho C \dot{T} - \rho T\frac{\partial^2 \psi}{\partial T \partial\gamma}\dot{\gamma},\label{ds}
\end{equation}
where $C = -T\frac{\partial^2\psi}{\partial T^2}\Bigr|_\gamma$ is also known as the heat capacity of the glass at constant strain. Considering now an elementary element of volume inside the sample, the heat equation per unit of volume element, that results from the first law of thermodynamics, is written as\cite{chrysochoos1989analyse}:
\begin{equation}
\rho\dot{e} = {\sigma} : {D} + r_{ext} - \vec{\nabla}\cdot \vec{q},\label{de}
\end{equation}
with ${D}$ being the strain rate tensor including the contribution of plasticity, $r_{ext}$ representing additional external heat sources if needed, and $\vec{\nabla}\cdot \vec{q}$ the heat exchanges at the surfaces. Combining Eqs. (\ref{rho}), (\ref{ds}), and (\ref{de}), the heat equation becomes:
\begin{equation}
\rho C\dot{T} + \vec{\nabla}\cdot \vec{q} = {\sigma} : {D} - \rho\frac{\partial\psi}{\partial\gamma}\Bigr|_T\dot{\gamma} + \rho T\frac{\partial^2\psi}{\partial T\partial\gamma}\dot{\gamma} + r_{ext}\label{eq:chaleur}
\end{equation}
\begin{figure*}
\includegraphics{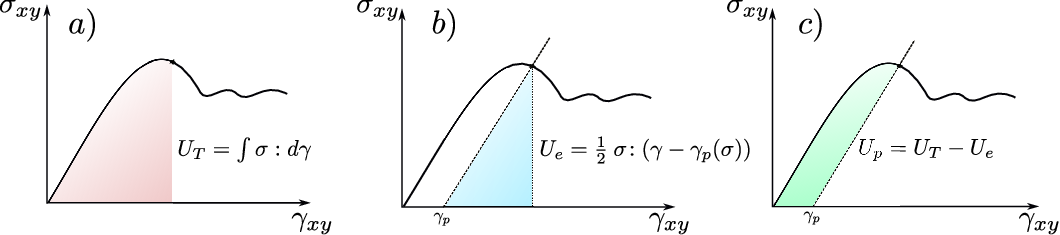}
\caption{\label{fig_3} Balance energy density as a function of time obtained as the respective area under the curve: $a)$ Total stored mechanical energy $U_T$. $b)$ Elastic contribution $U_e$. $c)$ Plastically dissipated energy $U_p$.}
\end{figure*}

\begin{figure*}
\includegraphics[width=\textwidth]{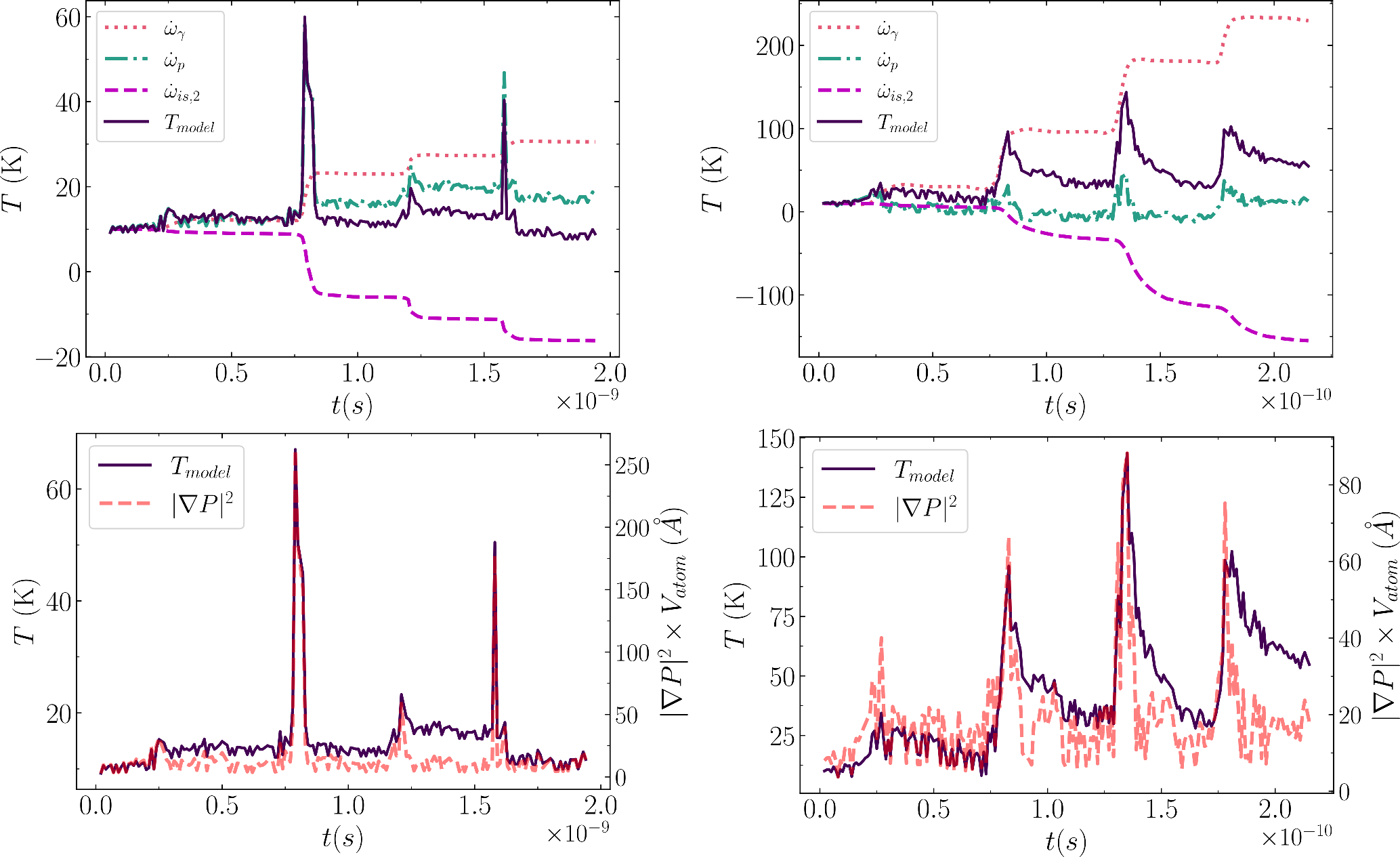}
\caption{\label{fig_4} (top) Atomistic calculation of the different contribution to the measured average temperature in SiO$_2$ (bottom) comparison between the temporal evolution of the average temperature, and the global average of the local quantity $\vert\nabla P\vert^2$ that contributes to gradient plasticity in SiO$_2$. (left) for $\dot\gamma=10^9$s$^{-1}$ (right) for $\dot\gamma=10^{10}$s$^{-1}$ }
\end{figure*}
\begin{figure*}
\includegraphics[width=\textwidth]{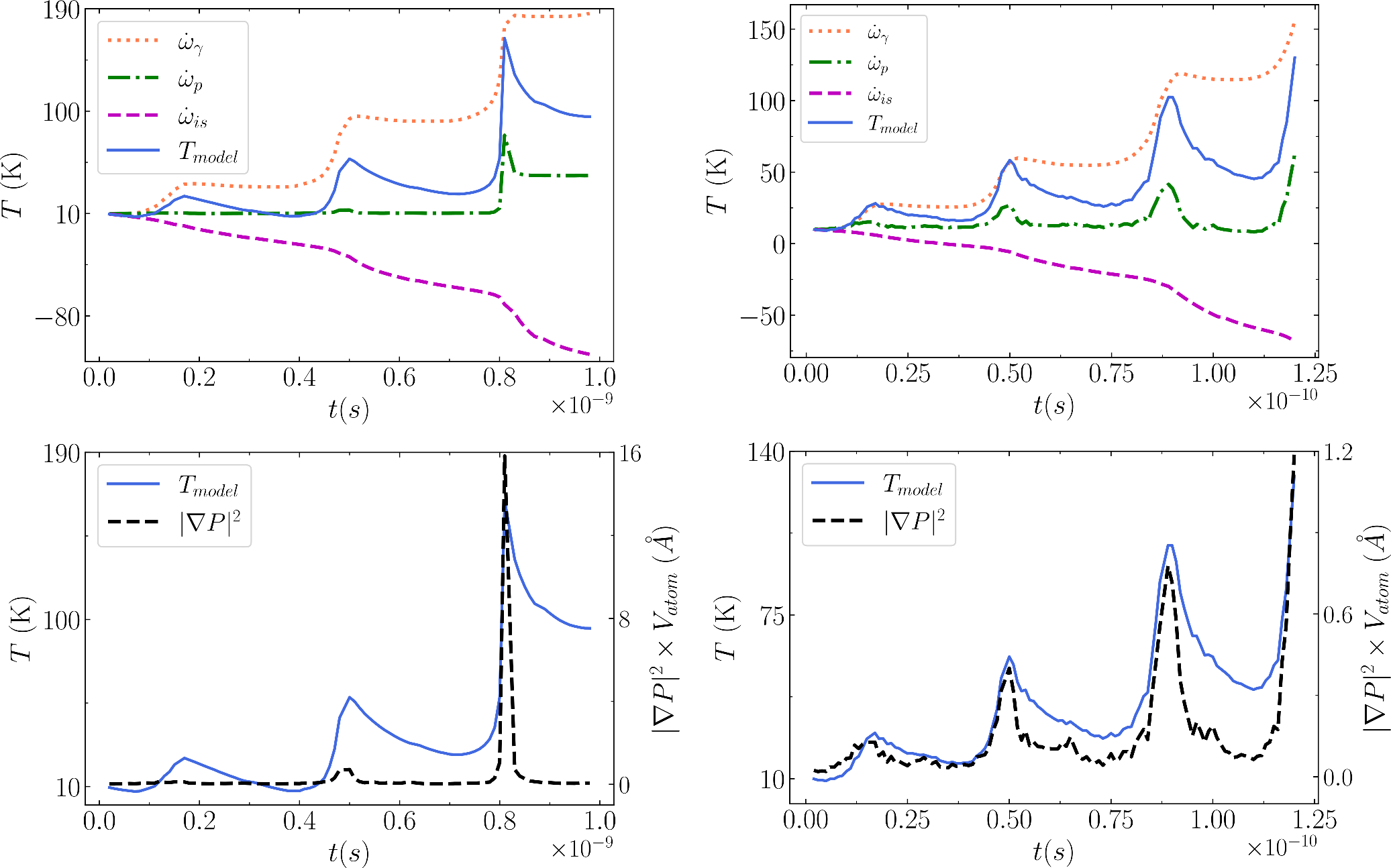}
\caption{\label{fig_5} (top) Atomistic calculation of the different contribution to the measured average temperature in CuZr (bottom) comparison between the temporal evolution of the average temperature, and the global average of the local quantity $\vert\nabla P\vert^2$ that contributes to gradient plasticity in CuZr. (left) for $\dot\gamma=10^9$s$^{-1}$ (right) for $\dot\gamma=10^{10}$s$^{-1}$ }
\end{figure*}

\begin{table*}[h!]
	\caption{\label{tab:table1}
		Fitting values for one-piece fit.}
	\begin{ruledtabular}
		\begin{tabular}{ccccccccc}
			Glass/Shear & $\dot{\gamma}$ ($s^{-1}$)  & $A_0$ (K$^{-1}$)  & $A_1 $ (K$^{-2}$) & $E$& $B$   & $l_p (\text{\AA})$ & $\mu$ (GPa) & $E_{rel}(\%)$\\
			\hline
			SiO$_{2}$ &$10^{9}$  & $2.5 \times 10^{-4}$ & $2.1\times 10^{-4}$ & $100$ & $0.02$& $8.4$ & $35.43$ & $23.51$\\
			
			&$10^{10}$  & $2.5\times 10^{-6}$ & $6.1 \times 10^{-7}$ & $100$ & $0.18$& $3.2$ & $35.54 $ & $23.07$\\
			\hline
			CuZr &$10^{9}$  & $6.0\times 10^{-3}$ & $0$ & $100$ & $0.2$ & $< 5.4 $ & $39.89 $ & $36.30$\\
			&$10^{10}$  & $2.5\times 10^{-5}$ & $0$ & $100$ & $0.14$ & $< 17 $ & $40.73 $ & $15.51$\\
			\hline \\
			\hline 
			Glass/Elongation & $\dot{\gamma}$ ($s^{-1}$)  & $A_0$ (K$^{-1}$)  & $A_1 $ (K$^{-2}$) & $E$& $B$   & $l_p (\text{\AA})$ & Young Modulus (GPa) & $E_{rel}(\%)$\\
			\hline
			CuZr &$10^{9}$  & $5.0 \times 10^{-6}$ &  $1.2 \times 10^{-4}$ & $100$ & $0.12$ & $25.5$ & $112$ & $14.22$\\
			&$10^{10}$  & $5.0\times 10^{-6}$ & $2.0\times 10^{-7}$ & $100$ & $0.11$ & $5 $ & $144.15$ & $15.54$\\
			\hline
		\end{tabular}
	\end{ruledtabular}
\end{table*}

\begin{table*}[h!]
	\caption{\label{tab:table2}
		Fitting values for two-part fit.}
	\begin{ruledtabular}
		\begin{tabular}{cccccccccccccc}
			Glass & $\dot{\gamma}$ ($s^{-1}$)  & $A0_1$(K$^{-1}$) & $A1_1$(K$^{-2}$) & $E_1$ & $B_1$ & $lp_1 (\text{\AA})$ & $Err_1 \%$ & $A0_2$(K$^{-1}$) & $A1_2$(K$^{-2}$) & $E_2$ & $B_2$ & $lp_2 (\text{\AA})$ & $Err_2 \%$ \\
			\hline
			SiO$_{2}$ &$10^{9}$  & $2.5\times10^{-4}$ & $2.1\times 10^{-4}$ & $100$ & $0.05$ & $8.4$ & $15.68$ & $2.5\times 10^{-4}$ & $2.1\times 10^{-4}$ & $100$ & $0.02$ & $9.2$ & $26.09$ \\
			
			&$10^{10}$  & $2.5\times 10^{-6}$ & $6.1\times^{-7}$ & $100$ & $0.18$ & $3.2$ & $17.63$ & $3.2\times 10^{-5}$ & $1.3\times 10^{-7}$ & $100$ & $0.12$ & $7.7$ & $27.31$ \\
			\hline
			CuZr &$10^{9}$  & $6\times 10^{-3}$ & $0$ & $100$ & $0.2$ & $0$ & $20.52$ & $1.2\times 10^{-2}$ & $0$ & $10$ & $0.26$ & $< 10$ & $22.67$ \\
			
			&$10^{10}$  & $2.5\times 10^{-5}$ & $0$ & $100$ & $0.14$ & $0$ & $10.24$ & $2.1\times 10^{-5}$ & $0$ & $100$ & $0.15$ & $< 13.4$ & $22.24$ \\
		\end{tabular}
	\end{ruledtabular}
\end{table*}

The first two terms on the RHS of Eq.~\ref{eq:chaleur} account for the intrinsic dissipation rate due to mechanical deformation and plasticity, which we will denote as $\dot{\omega}_d = \dot{\omega}_{\gamma} + \dot{\omega}_{p}$ with notations that will be explained later. We will consider that only a fraction $B$ of this dissipation is converted into heat, while the rest contributes, for example, to excite coherent waves. The third term accounts for the thermo-mechanical couplings and is commonly referred to as the isentropic heat rate $\dot{\omega}_{is}$ \cite{chrysochoos1992thermographic,sepulveda2024thermomechanical}. Finally, the last term $r_{ext} = 0$ since we are not considering external heat sources in our system. On the LHS at the global scale, following Chrysochoos\cite{chrysochoos1992thermographic}, the term $\vec{\nabla} \cdot \vec{q}$, i.e., the heat losses, can be simplified as a thermal attenuation contribution with a characteristic relaxation time $\tau$ which depends on the global thermal conductivity $\kappa$ as:
\begin{equation}
\tau = E\frac{\rho C L_y^2}{\kappa}
\label{Eq:tau}
\end{equation}
where $E$ is a proportionality constant to be determined. We can rewrite the heat equation now as:
\begin{equation}
\rho C \left(\dot{T} - \frac{T}{\tau}\right) = B(\dot{\omega}_{\gamma} + \dot{\omega}_{p}) + \dot{\omega}_{is}
\label{Heat}
\end{equation}
An analytical expression for the thermo-mechanical coupling term $\dot{\omega}_{is}$ is easily obtained by taking derivatives with respect to $\gamma$ and $T$ in Eq. (\ref{mech}). Thus,
\begin{equation}
\dot{\omega}_{is} = -2\mu \rho T \dot{\gamma}(A_0 + 2A_1(T-T_0))
\label{Eq:TherDil}
\end{equation}
The volumetric amount of heat generated per unit time by the conversion of dissipated mechanical energy (plastic work) $\dot{\omega}_d = \dot{\omega}_{\gamma} + \dot{\omega}_{p}$, into heat is computed as:
\begin{equation}
\dot{\omega}_\gamma = \frac{d}{dt}\left[ \int^{\gamma(t)}\sigma(\gamma) : d \gamma - \frac{1}{2} \sigma(t):\left( \gamma(t) -  \gamma_p( \sigma(t))\right)\right]\label{omega_gamma}
\end{equation}
as detailed in Fig.~\ref{fig_3}, and
\begin{equation}
\dot{\omega}_p = \frac{d}{dt}\left[\frac{2\mu l_p^2}{V}\int_V |\nabla p (t)|^2 d\Omega\right],
\label{omega_p}
\end{equation}
where in the framework of the strain gradient theory of plasticity $p$ is the equivalent atomic deformation defined around each atom $i$ supporting the atomic strain $\varepsilon_i$ as:
\begin{equation*}
\begin{split}
p_i = (\dfrac{1}{6}((\varepsilon_{i,xx}-\varepsilon_{i,yy})^2+(\varepsilon_{i,xx}-\varepsilon_{i,zz})^2 +  \\
(\varepsilon_{i,yy}-\varepsilon_{i,zz})^2) + ((\varepsilon_{i,xy})^2+(\varepsilon_{i,xz})^2+(\varepsilon_{i,yz})^2))^\frac{1}{2},\label{p_i}
\end{split}
\end{equation*}
$l_p$ is a material charateristic length to be determined, $V$ is the volume of the glass and $d\Omega$ is the elementary volume.  The contribution of $p$ to the dissipated energy is due to its spatial heterogeneities, as will be discussed later. In order to compute $\dot{\omega}_p$, we aproximate the integral in Eq~(\ref{omega_p}) by a sum over all the atoms in the glass:
\begin{equation}
\int_V|\vec{\nabla}p (t)|^2 d\Omega \approx \sum_i |\vec{\nabla}p_i(t)|^2 \Omega_i(t),\label{grad_p}
\end{equation}
where $\Omega_i(t)$ is the instantaneous Voronoi volume obtained by a Voronoi tesselation over $i^{th}$ particle at time $t$. To compute the equivalent atomic deformation $p_i$, we compute the atomic strain $\varepsilon_i$ from the infintesimal Cauchy strain tensor:
\begin{equation}
\varepsilon_{i,\alpha\beta} = \frac{1}{2}\left(\frac{\partial u_{\alpha}}{\partial\beta} + \frac{\partial u_{\beta}}{\partial\alpha} \right)_{\vec{r}=\vec{r}_i},\label{varepsilon}
\end{equation}
where $\alpha,\beta \in \{x,y,z\}$, $\vec{r}_i$ is the position of the $i^{th}$ particle and $\vec{u}$ is the coarse-grained continuous displacement field given by\citep{goldhirsch2002microscopic}
\begin{equation}
\vec{u}(\vec{r},t) = \frac{\sum_j m_j \vec{u}_j(t)\phi(\vec{r} - \vec{r}_j(t))}{\sum_j m_j\phi(\vec{r}-\vec{r}_j(t))},\label{uCG}
\end{equation}
with $\vec{u}_j(t)$ is the displacement of particle $j$, $m_j$ its mass, and $\phi(\vec{r})$ is the coarse-grained function which corresponds to a Gaussian function of width $\Delta_{cg}$. This continuous definition of the displacement allows preserving the mass and momentum conservation equation at all scales. We choose $\Delta_{cg}$ to be twice the nearest-neighbor interatomic distance, which, as reported in Ref. \cite{molnar2016sodium}, will avoid displacement and strain singularities in the coarse-grained quantities.

The second contribution $\omega_\gamma$ to the plastic work, is computed by substracting the elastic contribution from the total energy density, which is obtained by integrating the stress-strain curves shown in Fig.~\ref{fig_2}. Replacing $\Psi$ by $U_T$ in the calculation of $\omega_\gamma$, the elastic contribution to the total energy $U_e$ can be extracted from the reverse linear curve computed at each step $\gamma$. The plastically dissipated energy is then simply obtained as $U_p = U_T - U_e$. This procedure is illustrated schematically in Fig. \ref{fig_3}. Here, the total energy, represented by the pink area under the curve, is obtained by integrating the stress-strain curve $U_T = \int \sigma(\gamma) : d\gamma$ up to the current engineering strain $\gamma(t)$, its elastic contribution is the blue area of the triangle formed by the linear fit at each point of $\sigma$ intersecting at the corresponding intersect in $\gamma$, i.e. $\gamma_p(\sigma)$, thus $U_e=\frac{1}{2}\sigma (t) : (\gamma (t) - \gamma_p(\sigma))$. The plastically dissipated energy $U_p$, shown as the green area, is simply the difference between these two quantities. \textcolor{black}{The previously described method for computing plastic energy is valid only under the assumption that the system follows a linear backward constitutive law. In cases where this path is nonlinear, such as during simple elongation/compression, as illustrated in the Supplementary Material, we applied a full reverse loading cycle every 100 time steps. The dissipated energy per unit volume was then quantified as the area enclosed between the forward and the backward stress-strain curves obtained from these cycles.} Finally, we derive the plastically dissipated energy as a function of time to obtain: $\dot{\omega}_\gamma = \frac{dU_p}{dt}$.

With all the contributions now clearly defined, we can integrate Eq.~(\ref{Heat}) to obtain:
\begin{equation}
T(t) = T_0e^{t/\tau} + e^{t/\tau}\int_0^t \frac{1}{\rho C}e^{-u/\tau}(B(\dot{\omega}_\gamma + \dot{\omega}_p) + \dot{\omega}_{is}))du.
\label{Sol}
\end{equation}
This computed $T(t)$ will then be compared to the numerical measurement of $T$ in order to identify the underlying constitutive laws.

\section{Results}

\begin{figure*}
\includegraphics{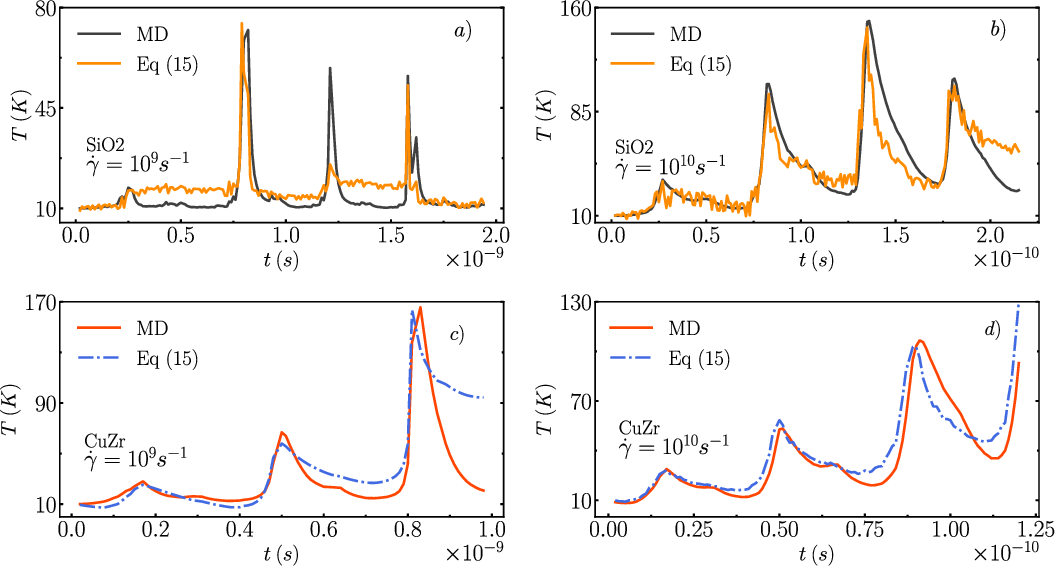}
\caption{\label{fig_6} Temperature evolution during the deformation shear cycles. $a)-b)$ SiO$_2$ at $\dot{\gamma} = 10^9 s^{-1}$ and $10^{10} s^{-1}$, respectively. The orange curve is the predicted temperature behavior obtained by Eq. (\ref{Sol}). $c)-d)$ CuZr at $\dot{\gamma} = 10^9 s^{-1}$ and $10^{10} s^{-1}$, respectively. The blue dashed curve is the predicted temperature behavior obtained by Eq. (\ref{Sol}).}
\end{figure*}

\begin{figure*}
\includegraphics{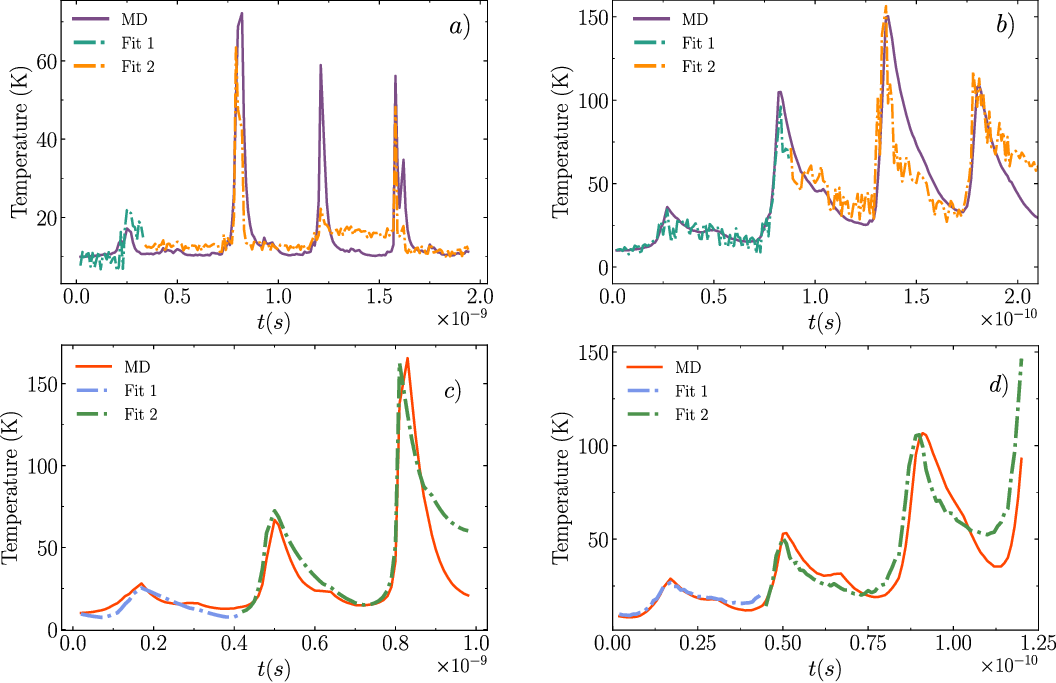}
\caption{\label{fig_7} Temperature evolution during the deformation shear cycles. $a)-b)$ SiO$_2$ at $\dot{\gamma} = 10^9 s^{-1}$ and $10^{10} s^{-1}$, respectively and $c)-d)$ CuZr at $\dot{\gamma} = 10^9 s^{-1}$ and $10^{10} s^{-1}$, respectively . Two parts fitting is used to improve the previous one part fit. Parameters are changing due to structural changes resulting from the plastic deformation taking place during the first part. The corresponding fitting parameters are regrouped in Table II.}
\end{figure*}

The different contributions to the temperature rises $T(t)$ are visible on Figures~\ref{fig_4} and~\ref{fig_5} for SiO$_2$ and CuZr, respectively. They all show the same kinds of behaviors but with different sensitivities. First, the conversion of dissipated energy into heat from local plastic deformation $\dot\omega_\gamma$: the time integral of this positive contribution induces an increasing temperature with constant stages during the elastic parts of the deformations decays. Second, the isentropic contribution resulting from the thermo-mechanical couplings included in the thermal dilatancy: this contribution is always negative and tends to decrease the temperature due to the related heat absorption. At low strain rates, this contribution may exceed the local plastic contribution, especially in SiO$_2$, thus preventing large temperature rises. A third contribution is however taken into account in the present model, due to non local plasticity as described in terms of strain gradient plasticity. This contribution results from spatial heterogeneities in the dissipation of energy: it induces large bumps of dissipated energy when shear bands occur. The temporal derivative (power) is positive before the occurrence of the shear bands, and then negative after the shear band occurred. Its integration results into temperature bursts related to the unfolding of extended spatial heterogeneities (see Figure~\ref{fig_8} for a visualization of these heterogeneities). As see in Figures~\ref{fig_4} and~\ref{fig_5}, these bursts coincide with the temperature peaks. However, they are very limited in the case of CuZr, but with very high amplitude in case of SiO$_2$, in agreement with the rapid temperature decay in this last case.
As can be seen in the numerical simulations, the relative contribution of these different terms to the global heating of the sample is strongly composition dependent. 

Once we determined $\dot{\omega}_{\gamma}, \dot{\omega}_{p}$, and $\dot{\omega}_{is}$, we used Eq.~(\ref{Sol}) to fit the temperature profile numerically computed via MD. The comparison between our model and the temperature tracked by MD is depicted in Fig.~\ref{fig_6}. We observe a noisy behavior in the temperature computed through Eq.~(\ref{Sol}). This noise arises from the computation of the Voronoi volume in $\dot{\omega}_p$. The bonds in SiO$_2$ tend to form rings, which have a significant effect on the Voronoi volume during deformation. In the metallic glass, this type of bonding is absent, and thus, we do not observe such noise. 

To fit the global temperature increase of the sample with only the three above mentioned energy contributions, we have introduced a set of six parameters and optimised the global agreement between the numerical temperature and the model on these set of parameters, including some physical constraints that defined the accessible range of values. The results are regrouped into Table I. and Table II.

The first parameter $E$ is an adjustment of the role of the boundary conditions on heat exchanges. It appears in the relaxation time $\tau$ as detailed in Eq.~(\ref{Eq:tau}). In general, we have identified $E=100$ that is larger that in the simplest approximate definition of $\tau$ given by the ration between the specific heat and the heat conductivity times the smallest accessible lengthscale $L_y$, but sometimes it is smaller: \textcolor{black}{in some cases, setting  $E = 10$ does not significantly alter the results, whereas $E = 1$ may no longer be appropriate. The physical interpretation of this parameter is however not very significant}, since the introduction of the relaxation times to describe heat exchanges at the boundaries is already an approximation. Interestingly, the order of magnitude of the relaxation time is that given by a uniform heat flux.

The second parameter $B$ represents the percentage of energy dissipated inside plastic zones that is converted into heat. The rest may be transformed into noise emission or electronic excitation for example. $B$ is necessarily smaller than $1$.  We have seen that it may depend on the strain rate, and on the composition as well (especially at small strain rates). 

$A_0$ and $A_1$ characterizes the thermal dilatancy and its temperature dependence, as described in Eq.~(\ref{Eq:TherDil}). Their units are $K^{-1}$ and $K^{-2}$ respectively. The contribution of $A_1$ is to amplify the temperature decay for high temperature levels after a heating. The orders of magnitude of their optimized values are close the that measured in real glasses\cite{gangopadhyay2020link}. However, it is interesting to note that $A_1=0$ during the \textcolor{black}{shear deformation cycles of the} CuZr metallic glass, indicating that no additional temperature decay is required following localized heating. \textcolor{black}{This behavior of the metallic glass does not hold during compression/elongation deformation where the glass behaves closer to the silica one. This may result from pronounced density fluctuations and bond breakage, which contribute to heat dissipation and temperature reduction.} More unattempted, $A_0$ is seen to decay with the strain rate \textcolor{black}{in all the samples}, may be due to the fact that at very large strain rates the system has not sufficient time to store energy through \textcolor{black}{thermal expansion. The effect is also visible in the first order coefficient $A_1$, thus denoting generally less sensitivity to the temperature at very high strain rates}. 

Finally, the last parameter $l_p$ is the characteristic size for the strain gradient contribution to the heat source, related to strain gradient plasticity and inhomogeneous deformation. \textcolor{black}{It is a nanometric size, but some specificities have to be mentioned on this typical size related to spatial heterogeneities. First, this size is well defined for SiO$_2$, but not for CuZr. This can be proved by looking at the error dependence on $l_p$ obtained by comparing the numerical fit to the measurement of the temperature as a function of time. This discrepancy shows a well-defined minimum in the $l_p$-space for SiO$_2$ submitted to shear, allowing to identify clearly a unique non-zero value of $l_p$. But for CuZr, the error bar is flat as a function of $l_p$ (see Fig. 6 in the supplementary material), thus allowing to identify only an upper bound. This means that $l_p=0$ is a sufficient choice in CuZr submitted to shear, proving that the spatial heterogeneities do not contribute significanlty to heat. However, when CuZr is submitted to an elongation, the situation is different. Again, a well defined value for $l_p$ is observed, and it appears to be larger (corresponding to a more homogeneous response in CuZr) than the values measured in SiO$_2$ case. This shows that strain heterogeneities become to play a role, under elongation, into the conversion of mechanical energy to heat. This can be related to the previous comment on the possible role of density fluctuations and bond breakage, as already discussed on the deformations type dependence of $A_1$. Another remarkable point concerns the strain rate sensitivity of $l_p$. In both cases: SiO$_2$ submitted to shear or CuZr submitted to an elongation, $l_p$ appears to decrease with the strain rate: from 25.5$\AA$ to 5$\AA$ in CuZr and from 8.4$\AA$ to 3.2$\AA$ in SiO$_2$ when the strain rate is increased from $10^9$ to $10^{10} s^{-1}$. This observation is in agreement with the observation that the shear bands become more serrated when the strain rate increases (see Fig.~\ref{fig_8} and~\cite{sepulveda2024thermomechanical}).} 

Table II. and Fig.~\ref{fig_7} \textcolor{black}{indicate clearly that all these parameters} may change during the deformation of the sample. This means that plastic deformation, and the related irreversible residual plastic strain, affect the structure of the glass and consequently its constitutive law. In general, the thermal dilatancy increases at larger strains, as well as the characteristic size for strain gradient plasticity, indicating a progressive enlargment of strain heterogeneities.

These results are very encouraging, because they prove that it is possible to find self-heating in glasses by using a simple constitutive law involving usual thermo-mechanical couplings based on thermal dilatancy. However it is remarkable to see that CuZr and SiO$_2$ have a very different behaviour. CuZr is a disordered assembly of metallic atoms linked with non-directional metallic bonds, while SiO$_2$ has an amorphous structure based on a distribution of small Si rings, with bridging Oxygens and iono-covalent directional Si-O bonding. The optimized fit shows, that the temperature increase is more peaked in SiO$_2$ and yields both additional contributions: first a  temperature dependence of the thermal dilatancy that contributes to decay the temperature proportionaly to its increase, second a strong contribution of non-local terms described by strain gradient plasticity, over nanometric lengthscales. In fact, the amplitude of the strain gradient is about one order of magnitude smaller in CuZr (Fig.~\ref{fig_5}) than it is in SiO$_2$ (Fig.~\ref{fig_4}). This article does not pretend to give an explanation to these differences, but its goal is to emphasize the strong composition dependence of the different terms in the constitutive law. We see clearly here that strain heterogeneities take place over smaller lengthscales in SiO$_2$, and contribute more as heat sources. But non-linear temperature dependence in the thermo-mechanical couplings contributes to reset the thermal increase more efficiently in SiO$_2$ than in CuZr.   

\begin{figure*}
\caption{\label{fig_8} Representative snapshots of the equivalent gradient deformation $p$ and the temperature $T$ during the shear deformation cycles. Capital letters (A,B,C) in the stress-strain curve indicate stress states for which the equivalent deformation $p$ and the temperature $T$ are shown. Only atoms with a temperature greater than $100$ K are depicted in the temperature snapshots. in the middle panel Strain is evaluated using a particle displacements within finite time intervals corresponding to 1$\%$ global deformation. On the bottom panel, the temperature profile is shown as a function of the orthogonal direction of formation of the shear bands $a)$ SiO$_2$ deformed at $\dot{\gamma}= 10^{10}s^{-1}$. $b)$ CuZr deformed at $\dot{\gamma}= 10^{10}s^{-1}$.}
\end{figure*}

To get more insights into the microscopic processes at play, we present in Fig.~\ref{fig_8} three stress states, denoted $A$,$B$ and $C$, of the SiO$_2$ in a), and of the CuZr glasses in b) during the deformation shear cycle at $\dot{\gamma} = 10^{10} s^{-1}$. 
For each glass, the snapshots in the middle panel show the deformation $p$ and temperature $T$ supported locally by the atoms in a representative plane inside the glass, during three illustrative states of the deformation. The bottom panel focuses on the local temperature supported by all atoms. As already proven in other amorphous samples~\cite{albaret2016mapping}, the plastic behavior is dominated by the appearance of isolated shear transformation zones (STZs) that start to appear prior to yielding. In SiO$_2$, the global yield for macroscopic plastic flow (maximum of global stress) occurs when the deformation is around $\gamma = 0.2$ [point $A$ in Fig.~\ref{fig_8}(a)]. At this point, we reverse the deformation until $\sigma_{xy} = 0$ and restart the deformation to generate more shear cycles. The second representative point was chosen during the shear softening region, at $\gamma = 0.35$ for both glasses. Here, the STZs turn into a system-spanning shear band [point $B$ in Fig.~\ref{fig_8}(a)]. Finally, our last representative stress state is chosen when we finish the shear cycle at $\gamma = 0.4$ [point $C$ in Fig.~\ref{fig_8}(a)]. The same characteristic points have been detailed for CuZr in Fig.~\ref{fig_8}(b). In order to examine the contribution to the temperature in the shear band (SB), we compute the local temperature of each atom $(i)$ from its kinetic energy using $2 K(i)/(3 k_B) = T(i)$, where $K(i)$ is the kinetic energy of the atoms $i$ (after having removed the global shear velocity) and $k_B$ is the Boltzmann constant. The temperature profile in the bottom panels is shown perpendicularly to the shear band direction (either as a function of the $y$-coordinate - that is along the transverse direction of the global shear deformation, or as a function of the $x$-coordinate when the shear band unfolded along the $y$ axis). These observations, made at the atomic scale on both glasses, support the previous assertion that the atoms that contribute the most to the self-heating are those which have a larger equivalent strain $p$. To highlight this effect in the temperature snapshots, we only include atoms with $T(i) > 100$ K. We do not include the relevant snapshots of the alloys at $\dot{\gamma} = 10^{9} s^{-1}$ to avoid repetition. However, we include them in the \textcolor{black}{supplementary material, along with a structural analysis based on the static structure factor (pair distribution function) that shows that the small number of cycles studied here, although sufficient to trigger strain localization, is not sufficient to change the atomic structure significantly. The same remark holds for the potential energy. However, the distribution of the non-affine displacement field measured with a 0.1$\%$ strain step applied at different stages of the deformation shows a monotonous increase of the average amplitude of the non-affine displacement field  with the number of cycles applied, without affecting its distribution shape. As a reminder, the non-affine displacement field is obtained by substracting to the total displacement supported by each atom the affine displacement that would result from the average homogeneous strain~\cite{Tanguy2002} (an interesting quantity that has the advantage to be tractable in perturbative calculations). The results shown here  mean that the global characteristics (distribution) of the non-affine displacements are not affected by the deformation, while their amplitude (dynamics) can be strongly history dependent, even when not any significant structural change is observed.}. 

We show in the bottom panels, that the temperature at the atomic scale may reach few thousands Kelvin, especially when the plastic deformation turns into a system-spanning shear banding. This has already been observed in experimental systems~\cite{Greer2013}, while the exact amplitude of the temperature increase seems to depend strongly on the composition of the glass~\cite{Greer2006,Wright2009} as well as on the strength of the deformation~\cite{Ketov2013}. Interestingly, it has also been observed~\cite{Greer2013}, as in the present work, that the size of the thermally affected zones is far larger that the size of the shear bands. In our case, the observed shear band has a nanometric size very close to the measured $l_p$, while the thermal increase is centered on the shear band, but spans over larger sizes, with a linear decay down to the thermalized boundaries of the box. The occurrence of shear bands is thus always related to a local thermal increase, but with a larger extent than that of the spatial strain heterogeneity.

\section{Conclusion}

To conclude, we have shown using molecular dynamics simulations based on the numerical resolution of the classical equations of motion, and a continuous modelization of the thermo-mechanical constitutive laws, that self-heating in two different compositions of glasses can be described by considering the irreversible plastic deformation as a heat source. In SiO$_2$ glasses, the plastic activity involves non local terms (described as strain gradient plasticity) due to the strong signature of strain localization in the thermal increase in these glasses. The strong thermal increase is counterbalanced by a non-linear temperature dependence in the thermo-mechanical couplings. The thermal dilatancy is indeed mainly reponsible for the temperature decay after the thermal peaks. In CuZr, these two terms are not necessary, a simple description based on the analysis of the hysteretic behaviour in the stress-strain relationship and on a constant thermal dilatancy is sufficient to recover the right thermal increase induced by the cyclic mechanical deformation. The thermal dilatancy, while rarely taken into account in the constitutive laws of glassy samples is thus very important to recover their right thermo-mechanical behavior.
It is shown, in this case, that the local thermal increase may even overcome the melting temperature, especially during catastrophic nucleation of system spanning shear bands. The global average temperature remains however in limited ranges.

The comparison between two different strain rates shows that the thermal expansion and the size of the strain heterogeneities may be strain rate dependent.  \textcolor{black}{In general, the thermal expansion appears to decrease with the strain rate, as well as the characteristic size of the strain heterogeneities, as already observed long time ago in amorphous silicon~\cite{fusco2010role}. However, the true values of the strain rates accessible with molecular dynamics simulations (very high strain rates) are not sufficiently understood to allow a realistic comparison with experimental results, so only qualitative trends can be retained. It is also important to remind, as already mentioned in our previous work~\cite{sepulveda2024thermomechanical}, that finite size effects affect strongly the results in the plastic regime, but they are out of scope of the present paper.}

Finally, this study emphasizes the importance of non-local terms used in strain gradient plasticity to take account of the intermittency in the heterogeneous plastic flow as a function of strain. This heterogeneous flow acts as a heat source over nanometric lengthscales, but contributes as well to thermal decrease when the strain gradient becomes smaller. This contribution appears especially significant in \textcolor{black}{systems submitted to an elongation, and in SiO$_2$ glasses in which bond directionality acts certainly in favor of strain localization and density fluctuations on small } lengthscales.

\begin{acknowledgments}
The authors are grateful to Djimedo Kondo for fruitful discussions. They acknowledge Gergely Molnar for drawing our attention to theories of gradient plasticity that show the role of spatial heterogeneities as an additional source of time-limited dissipation.
This work was supported by the RATES project (ANR-20-CE08-0022) for Matias Sepulveda and Anne Tanguy, and by the DENSE project (ANR-21-CE08-0005) for Rene Alvarez-Donado and Anne Tanguy. Both projects are founded by the French National Research Agency. Finally, we wish to acknowledge the support of the author community in using
REV\TeX{}.
\end{acknowledgments}

Data Availability: The data that support the findings of this article are openly available in ~\cite{Data-2025}.

\end{document}